\documentclass[a4paper,11pt]{article}
\pdfoutput=1 % if your are submitting a pdflatex (i.e. if you have
\usepackage{jheppub} % for details on the use of the package, please
\usepackage[T1]{fontenc} % if needed

\usepackage{graphicx}% Include figure files
\usepackage{dcolumn}% Align table columns on decimal point
\usepackage{bm}% bold math
\usepackage{savesym}
\usepackage{amsmath, amsfonts, amsthm, amssymb}
\savesymbol{iint}
\usepackage{txfonts} % this is the font
\restoresymbol{TXF}{iint}
\usepackage[ansinew]{inputenc}
\usepackage{textcomp}

\title{\boldmath  Scalar Dark Matter Candidates in Two Inert Higgs Doublet Model}

\author[a]{E. C. F. S. Fortes,}
\author[a,1]{A. C. B. Machado,\note{Corresponding author.}}
\author[a]{J. Monta\~no,}
\author[a]{V. Pleitez}

\affiliation[a]{Instituto de F\'\i sica Te\'orica, Universidade Estadual Paulista, Rua Dr. Bento Teobaldo Ferraz 271,
01140-070 S\~ao Paulo, SP, Brazil}

\emailAdd{elaine@ift.unesp.br}
\emailAdd{ana@ift.unesp.br}
\emailAdd{montano@ift.unesp.br}
\emailAdd{vicente@ift.unesp.br}

\abstract{We study a two scalar inert doublet model (IDMS$_3$) which is stabilized by a $S_3$ symmetry. We consider two scenarios: i) two of the scalars in each charged sector are mass degenerated due to a residual $\mathbb{Z}_2$ symmetry, ii) there is no mass degeneracy because of the introduction of soft terms that break the $\mathbb{Z}_2$ symmetry. We show that both scenarios provide good  dark matter candidates for some range of parameters.}

\begin{document}
\maketitle
\flushbottom

\section{Introduction \label{sec1}}

The existene of dark matter (DM) has been well established since the early astronomical~\cite{Rubin:1970zza} and cosmological observations~\cite{Riess:1998dv,Perlmutter:1998np,Bahcall:1999xn,Larson:2010gs}. For more recent data see \cite{Beringer:1900zz}. It accounts for approximatively $23 \%$ of the composition of the universe. Moreover,  these observational evidence justify the
experimental searches trying to find events that can be interpreted as direct manifestations of DM. Some of them are astronomical observations~\cite{Cline:2014fua, LH11, HL11,Hooper:2007kb,Hooper:2010mq}, and others like DAMA~\cite{Bea10},
CoGeNT~\cite{Aea111}, CDMS~\cite{Aea211}, XENON~\cite{Aea311,Aea411}, and LUX~\cite{Akerib} are experiments trying to measure the recoil energy of nuclei if it scatters with the DM.

Models which contain DM candidates have to explain among other aspects, the DM density, which is $\Omega h^2 =
\rho h^2/\rho_c = 0.1196 \pm 0.0031$ where, $h$ is the scale factor for Hubble expansion \cite{Planck}, $\rho_c = 3H_0^2/\left(8\pi G\right)$ is the critical density of the Universe, and $H_0$ is the current value of the Hubble constant~\cite{Beringer:1900zz}.

Much effort has been employed in order to discover or interpreted DM signals. It is possible that it consists of one or more elementary particles which interact very weakly with ordinary matter. One of the most common scenarios are supersymmetric models~\cite{Bertone:2004pz}. In fact, in this kind of models the lightest supersymmetric particle (neutralino) is prevented by the $R$ parity to interact with the known particles. The neutralino is an exemple of cold dark matter (CDM), i.e. a kind of DM which is not relativistic at the time of freese out. Of course, there are other possibilities, for instance, Kaluza-Klein states in models with universal~\cite{Kolb, Servant} or warped~\cite{Agashe} extra dimensions, stable states in little Higgs theories \cite{Birkedal} and a number of models with extra heavy neutrinos.  Some other alternative scenarios for DM consider self-interacting DM \cite{Fortes:2013ysa} and warm DM \cite{ZV12}. Other ambitious scenarios  consider asymmetric dark matter models. They have their motivation based on the similarity of mass densities of the DM $(\rho_{DM})$ and that of the visible matter $(\rho_B)$ observed $\rho_{DM}/\rho_{B}\approx5$ and try to explain this rate. Consequently, most of these models are based on the hypothesis that the present abundance of DM and visible matter have the same origin~\cite{Zurek:2013wia,Petraki:2013wwa}.

An additional  and interesting scenario which contains DM candidates is the inert doublet model
(IDM)~\cite{LopezHonorez:2006gr,Hambye:2009pw,LopezHonorez:2010tb,Honorez:2010re,Gustafsson:2012aj}. It is a minimal extension of
the SM  which contains a second Higgs doublet ($H_2$) with no direct couplings to quarks and leptons. The first time that the phenomenology
of an inert doublet  was considered was in the context of neutrino physics~\cite{Ma:2006km} and also in the context of the problem of
naturalness~\cite{Barbieri:2006dq}. In all cases, the inert doublet was possible due to a $\mathbb{Z}_2$ symmetry under which $H_2\to - H_2$ and all the other fields are even. In particular, this discrete symmetry forbids interactions like
$(H^\dagger_1H_2)(H^\dagger_2H_2)$, being $H_1$ the SM Higgs doublet.

In this work, we study the three Higgs doublet model with a $S_{3}$ symmetry, proposed in ~\cite {Machado:2012ed}, in which,  besides the standard model-like doublet there are two additional inert doublets, here denoted $H_2$ and $H_3$. It is this $S_3$ symmetry and an appropriate vacuum alignment that allows us to obtain a model with two inerts doublets. Besides, we already know from IDM models that this new particles have a rich phenomenology, especially as a good dark matter candidate. Here we will show that the same can happens in a model with two inert doublets.  We will analyze two scenarios for this model, one
in which the extra scalars are mass degenerated and the other in which soft terms, breaking a residual $\mathbb{Z}_2$ symmetry, are added, resulting in non degenerated masses for this extra scalars.

The paper is organized as follows. In Sec.~\ref{sec:2} we briefly present the model. In Sec.~\ref{sec:3} we briefly describe the theoretical framework for the calculations for DM abundance and in the Sec.~\ref{sec:4} we show the parameter choices suitable for the dark matter candidate and the numerical results. Finally in the
last section section, Sec.~\ref{sec:5}, we summarize our conclusions.

\section{The Model}
\label{sec:2}

In the context of standard model (SM)  the number of scalar doublets can be arbitrary. An interest case is that the number of these fields is the same as the number of the fermion families, i.e. just three. In this case, as we said before, the $S_3$ symmetry is, probably, the most interesting one because it is the minimal non-abelian discrete symmetry with one doublet and one singlet irreducible representations.

The model that we will consider here has the three Higgs doublets transforming as $(\textbf{2},+1)$ under $SU(2)_L\otimes U(1)_Y$ and under $S_3$ as:
\begin{eqnarray}
&&S=H_1\sim\textbf{1},\nonumber \\&&
(D_1,D_2)=\left(H_2, H_3\right)\sim\textbf{2}.
\label{ma}
\end{eqnarray}
This case was called Case B in Ref.~\cite{Machado:2012ed} and we will be restricted to this case in the present paper. The necessary conditions under which the vacuum alignment $v_1=v_{SM}$, $v_{SM}$ is the SM VEV $\sim246$ GeV and $v_2=v_3 = 0$, allow a scalar potential bounded from below and a stable minimum as has been shown in Ref.~\cite{Machado:2012ed}. With this vacuum alignment and, since the quarks and leptons are singlet of $S_3$,  the two Higgs doublet $D_1,D_2$ do not couple to fermions and do not contribute to the spontaneous symmetry breakdown, i.e they are inerts. They couple only to the gauge bosons and this vacuum alignment also implies in a residual $\mathbb{Z}_2$ symmetry in which the two inert doublets are mass degenerate in each charged sector. In this case the mass spectra is
\begin{eqnarray}
&&m_{H_{2}^{0}}^{2}=m_{H_{3}^{0}}^{2}=\mu_{d}^{2}+\frac{1}{2}\lambda^{\prime}v_{SM}^{2},
\nonumber \\&&
m_{A_{2}}^{2}=m_{A_{3}}^{2}=\mu_{d}^{2}+\frac{1}{2}\lambda^{\prime\prime}v_{SM}^{2},
\nonumber \\&&
 m_{h_{2}^{+}}^{2}=m_{h_{3}^{+}}^{2}=\frac{1}{4}(2\mu_{d}^{2}+\lambda_{5}v_{SM}^{2}),
\label{massas1}
\end{eqnarray}
where $\mu_d^2$ came from the term $\mu_d^2 [D^\dagger D]_1$ in the scalar potential, with $\lambda^{\prime}=\lambda_5+\lambda_6+2\lambda_7$ and $\lambda^{\prime\prime}=\lambda_5+\lambda_6-2\lambda_7$,
with $\lambda_{5,6,7}$ are quartic coupling constants in the scalar potential. We call this Scenario~1.

If the residual $\mathbb{Z}_2$ symmetry is softly broken by adding non-diagonal quadratic terms in the inert sector, the mass degeneracy is broken and the mass spectra becomes
\begin{eqnarray}
&&\overline{ m}_{H_{2}^{0}}^{2}=m_{H_{2}^{0}}^{2} - \nu^{2},\quad
\overline{ m}_{H_{3}^{0}}^{2}=m_{H_{3}^{0}}^{2}+\nu^{2},
\nonumber \\   &&
\overline{m}_{A_{2}}^{2}=m_{A_{2}}^{2} - \nu^{2},\quad
\overline{m}_{A_{3}}^{2}=m_{A_{3}}^{2}+\nu^{2},
\nonumber\\&&
\overline{m}_{h_{2}^{+}}^{2}=m_{h_{2}^{+}}^{2} - \nu^{2},
\quad
\overline{m}_{h_{2}^{+}}^{2}=m_{h_{2}^{+}}^{2}+\nu^{2},
\label{massas2}
\end{eqnarray}
and we call this Scenario 2.

In the case of mass degenerate scalars, the lightest scalars can be DM candidates and we will choose the $CP$ even ones. In the case of no mass degeneracy it is possible that the lightest one is the DM candidate. For the Scenario 1, our parameter choice enables us to establish the follow order for the mass of the scalars: $ m_{A_{2,3}}> m_{h_{2,3}^{+}}> m_{H_{2,3}^{0}}$. Since $H^0_{2,3}$ are the lightest neutral scalars, their decays are kinematically forbidden.  With the rearrangement of the parameters, instead of choosing $H_{2,3}^{0}$, we could choose the $CP$ odd
scalars $A_{2,3}$ as the DM candidates, if they were  the lightest ones, and the same conclusions would
remain valid for this scenario. In the Scenario 2, $H_{2}^{0}$ accounts for all the $\Omega_{DM}h^2$ contribution. In each scenario we choose two
set of parameters as is shown in Table~\ref{tab1}.

\section{Dark Matter Abundance}
\label{sec:3}

Preliminary analysis showing that this model can accommodate dark matter candidates were done in
Ref.~\cite {Machado:2012ed}. Here, this will be confirmed by a more detailed analysis. In order to calculate the DM
abundance we have used the MicrOMEGAs package to solve numerically the Boltzmann equation after implementing all the
interactions of the model in the CalcHEP package~\cite{Belyaev}. \par

Let us consider for instance, the model of inert doublets with non degenerated mass (Scenario 2). In this case,  as
we already said in Sec.~\ref{sec:2},  $H^0_{2}$ is our DM candidate. The evolution of the numerical density
$n$ of $H_{2}^{0}$, at the temperature $T$ in the early Universe, is given by the Boltzmann equation , which is written in simplified
form as follows~\cite{Belanger:2010pz}:
\begin{equation}
\frac{dY}{dy} = -\sqrt{\frac{\pi
g_*}{45G}}\frac{m_{H_2^{0}}}{y^2}\langle\sigma_{\rm ann}|\textrm{v}|\rangle\left(Y^2 -
Y_{eq}^2\right),
\label{bol}\end{equation}
here  $Y = n/s$, $s$ is the entropy per unity of volume, $Y_{eq}$ is the $Y$ value in the thermal equilibrium,
$y = m_{H^0_{2}}/T$. The parameter $G$ in Eq.~(\ref{bol}) is the Newton gravitacional constant, $\sigma_{\rm ann}$
is the cross section for annihilation of the particle $H_2^{0}$ and $\textrm{v}$
is the relative velocity, and the symbol $\langle\rangle$ represents thermal average. Finally, $g_*$ is a
parameter that measures the effective number of degrees of freedom at freeze-out, which is expressed as
\begin{equation}\label{aa}
g_{*}=\sum_{i=bosons}g_{i}\left(\frac{T_{i}}{T}\right)^{4} +\frac{7}{8}\sum_{i=fermions}g_{i}
\left(\frac{T_{i}}{T}\right)^{4},
\end{equation}
where the sums runs over only those species with mass $m_{H_{2}^{0}}\ll T$~\cite{KT98}. The model studied here has, besides the SM particles, 8 extra scalars ($A_{2}^{0}, A_{3}^{0},H_2^{0},H_3^{0},h_2^{\pm},h_3^{\pm} $). So, considering, for instance $T\gtrsim 300$ GeV we obtain $g_*\approx 114.75$.

To find $Y_0$, the present value of $Y$,  Eq. (\ref{bol}) must be integrated between $y = 0$ and $y_0 = m_{H_2^{0}}/T_0$.
Once this value is
found, the contribution of $H_2^{0}$ to DM density is
\begin{equation}
\Omega_{h_2} = \frac{m_{H_{2}^{0}} s_0Y_0}{\rho_c}.
\end{equation}
The same calculations hold for the Scenario 1.

\section{Results and Comments}
\label{sec:4}

The main numerical results for this model are presented in this section. We present in Table~\ref{tab1} the parameters
choice for both scenarios. The interactions and Feynman rules can be found in Ref.~\cite{Fortes}. For both scenarios (1 and 2), we have considered some set of parameters, so we call these scenarios respectively scenario 1a, 1b and scenario 2a, 2b and 2c. 

For scenario 1a, the dominant annihilation channels are: 39\% relative to $H_{3}^{0}H_{3}^{0} \rightarrow b\overline{b}$, 39\%  to $H_{2}^{0}H_{2}^{0}\rightarrow b\overline{b}$, 5\%  to
$H_{3}^{0}H_{3}^{0}\rightarrow GG$, 5\% to $H_{2}^{0}H_{2}^{0}\rightarrow G G$, 4\%  to $H_{3}^{0}H_{3}^{0}
\rightarrow \tau^{+}\tau^{-}$, 4\%  to $H_{2}^{0}H_{2}^{0}\rightarrow \tau^{+}\tau^{-}$, 2\% to $H_{3}^{0}H_{3}^{0}\rightarrow c\overline{c}$ and 2\% due to $H_{2}^{0}H_{2}^{0}\rightarrow c\overline{c}$. The
contribution of the two candidates ($H_{3}^{0},H_{2}^{0}$) to the Higgs invisible decay is 34.8\%. The Higgs invisible decay depends strongly on the parameter $\lambda^{\prime}$. In scenario 1, another choice of
parameters which brings null contributions to this invisible decay is reached with the numbers presented in the scenario 1b. The dominant annihilation channels are in this case 50\% relative to $H_{3}^{0}H_{3}^{0}\rightarrow W^{+}W^{-}$ and 50\% relative to $H_{2}^{0}H_{2}^{0}\rightarrow W^{+}W^{-}$.

Next we consider Scenario 2, in which $H_{2}^{0}$ is the only DM candidate. In Scenario 2a,  the dominant annihilation channels are respectively 77\%  relative to $H_{2}^{0}H_{2}^{0}\rightarrow b\overline{b}$, 11\% to $H_{2}^{0}H_{2}^{0}\rightarrow GG$, 8\% to $H_{2}^{0}H_{2}^{0} \rightarrow \tau^{+}\tau^{-}$ and 3\% due to $H_{2}^{0}H_{2}^{0}\rightarrow c\overline{c}$. In this scenario, $H_{2}^{0}$ doesn't contribute to the Higgs invisible decay.

For all the scenarios discussed above, the Cold DM-nucleons amplitudes are in agreement with CoGent, DAMA, LUX, XENON100. The Scenario 2b and 2c are in agreement with the predictions of XENON1T for $\sigma^{SI}$. In scenario 2b, the dominant annihilation channels are 78\% relative to $H_{2}^{0}H_{2}^{0}\rightarrow b \overline{b}$, 10\% relative to $H_{2}^{0}H_{2}^{0}\rightarrow GG$, 8\% relative to $H_{2}^{0}H_{2}^{0}\rightarrow \tau^{+}\tau^{-}$ and 3\% relative to $H_{2}^{0}H_{2}^{0}\rightarrow c\overline{c}$. In scenario 2c, the dominant annihilation channel is 100\% due to  $H_{2}^{0}H_{2}^{0}\rightarrow W^{+}W^{-}$. In this scenario, a negative $\lambda_{5}$ favors mainly the Higgs decay into two neutral gauge bosons \cite{Fortes}. Due to the smallness of $\lambda^{\prime}=0.001$, in scenario 2b the branching $h\rightarrow H_{2}^{0}H_{2}^{0}$ is negligible ($\approx 5 \times 10^{-4}$), since this Higgs decay is very sensible to this parameter. 

The Fig.~\ref{exp} shows the data presented in Table\ref{tab1} compared to the experimental results for $\sigma^{SI}$ considered in the experiments CoGent, DAMA, LUX, XENON100 and XENON1T.

% % % % % % % % % % % % % % %

\section{Conclusion}
\label{sec:5}

Here we have considered a two inert doublet model with an $S_3$ symmetry. The model has, besides the SM particles, eight scalars bosons which are inert, i.e.  they do not contribute to the spontaneous electroweak symmetry breaking.  They interact only with the gauge bosons through trilinear and quartic interactions, here only the latter one is important. In the case of degenerated masses (Scenario 1), two neutral scalars plays the  role of DM and in the case of non-degenerated masses (Scenario 2), one of the neutral scalars is the DM candidate. Besides these candidates, depending on the parameter choice, the model can also accommodate pseudoscalars DM candidates. It is well known that in the one inert doublet model there exist a set of allowed parameters in which we have a dark matter candidate and, in particular, that  there are three allowed regions of masses that are compatible with observed value of $\Omega_{DM}h^2$ and $R_{\gamma\gamma}$: i) $\lesssim$ 10 GeV; ii) 40-150 GeV, and iii) $\gtrsim$ 500 GeV~\cite{Krawczyk:2013jta}. Here we have proved that the IDMS$_3$ also has DM candidates at least in the second region, the analysis of the other regions will be considered elsewhere.

We have analyzed, as an illustration, some possible set of  parameters in both scenarios for the scalars contributions to $\Omega_{DM} h^2$.  We
call them Scenario 1a, 1b, 2a, 2b and 2c. It can be seen from the Table and the figure that the spin-independent elastic cross
section, $\sigma^{SI}$, is in good agreement with the results of experiments LUX and XENON100 for the mass range of DM considered here. We also have presented  scenarios (2b and 2c) where the predictions of XENON1T, to be measured in the future, are matched. The cross section $\sigma^{SI}$, as can be seen from the Table\ref{tab1}, are strongly dependent of the parameter $\lambda^{\prime}$.

The contribution to the ratio $R_{\gamma\gamma}$ in the present model has interesting differences compared to one inert doublet and that will be shown
elsewhere~\cite{Fortes}.

\acknowledgments

ACBM thanks  CAPES for financial support. ECFSF and JM thanks to FAPESP under the respective processes numbers 2011/21945-8 and 2013/09173-5. VP thanks CNPq  for partial support. One of us (ECFSF) would like to thank A. Pukhov for helpful discussions.

\newpage
\begin{table}[ht]
  \centering
\begin{tabular}{|c||c|c|c|c|c|}
\hline
    & scenario 1a &scenario 1b &scenario 2a&scenario 2b&scenario 2c\\
   \hline \hline
 $m_{H_{2}^{0}}$ & 54.1&79.9 & 63.4 & 59.1& 168\\
 $m_{H_{3}^{0}}$ & 54.1&79.9 & 86.59 & 83.47& 178.04\\
 $m_{A_{2}^{0}}$ & 112.44&127.95 &117.19 & 117.25& 196.16\\
 $m_{A_{3}^{0}}$ & 112.44&127.95  & 131.19 &131.24 & 204.83\\
 $m_{h_{2}^{+}}$ & 85.02&95.36 & 83.09 &83.13 & 84.70\\
 $m_{h_{3}^{+}}$ & 85.02&95.36 &101.89 &101.92 & 103.21\\
 $\mu_{d}$ & 48.53 &78.1& 72 &72.1 & 173\\
 $\nu$ & $-$ &  $-$&41.7 &41.7 & 41.7\\
 $\lambda^{\prime}$ & 0.019 &0.009& 0.019 &0.001 & 0.001\\
   $\Omega$ & $0.11$& 0.11&0.108 &0.11 & 0.11\\
 $\sigma v$ &$0.0832$ & $0.003$& $6.17$ &$0.0013$ & 0.74\\
  $\sigma^{SI}_{proton}$ &$7.33\times 10^{-46}$& $7.44\times 10^{-47}$ & $5.31\times 10^{-46}$ & $1.7\times 10^{-48}$ & $2.019\times 10^{-49}$\\
  $\sigma^{SI}_{neutron}$ &$8.38\times 10^{-46}$ &$8.52\times 10^{-47}$ & $6.08\times 10^{-46}$&$1.9\times 10^{-48}$ & $2.32\times 10^{-49}$\\
\hline
\end{tabular}
\caption{Parameters choice for Scenario 1 and 2 with $m_h=125$ GeV. The other masses  units are in GeV, $\sigma v$ is in units of $10^{-26}$ $cm^{3}/s$ and the units for  $\sigma^{SI}$ are in $cm^{2}$. The parameters $\lambda^{\prime\prime}=0.34$ and $\lambda_{5}=0.4$ for  scenarios 1a,1b, 2a, 2b and $\lambda_{5}=-0.4$ for scenario 2c.} \label{tab1}
\end{table}

\newpage

{\begin{figure}[ht]
 \begin{center}
  \includegraphics[width=15 cm, height=10 cm]{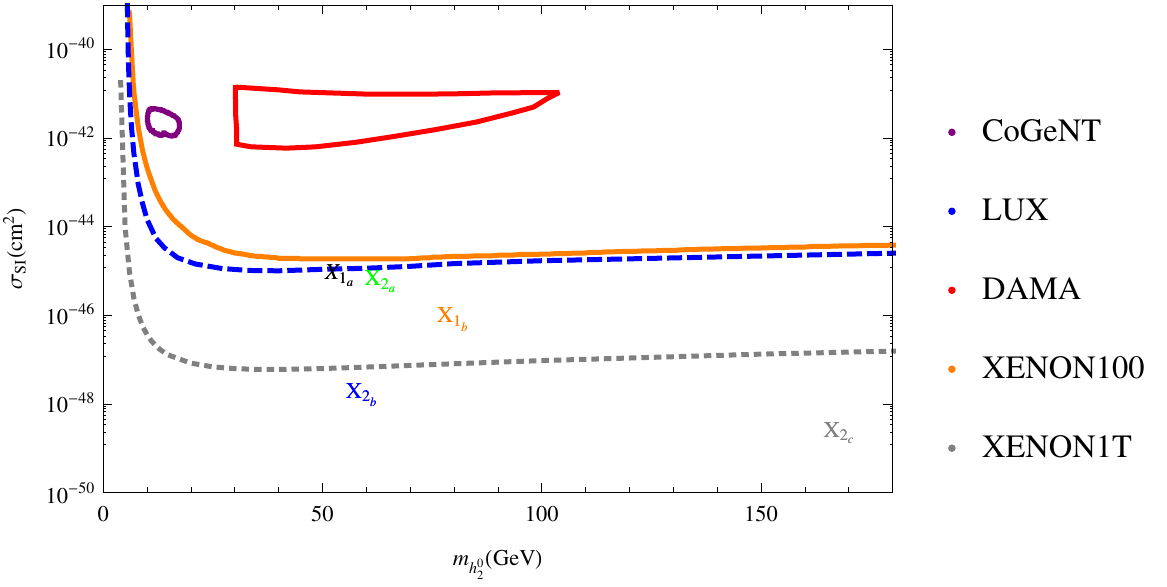}
   \caption{Limits for $\sigma_{SI}$ according to the experiments CoGent, DAMA, XENON100, XENON1T and LUX.\label{exp}. The points $X_{1a}$, $X_{1b}$, $X_{2a}$, $X_{2b}$ and $X_{2c}$ are the ones refer to scenarios $1a$, $1b$, $2a$, $2b$ and $2c$ given in Table\ref{tab1}.}
\end{center}
\end{figure}}

\end{document}